\documentclass[12pt]{article}
\usepackage{epsfig,psfig,float,amssymb,latexsym}
\usepackage[notref,notcite]{}
\newcommand{\NP}[1]{ Nucl.\ Phys.\ {\bf #1}}
\newcommand{\PL}[1]{ Phys.\ Lett.\ {\bf #1}}
\newcommand{\PR}[1]{Phys.\ Rev.\ {\bf #1}}
\newcommand{\bi}{\bibitem}
\newcommand{\be}{\begin{equation}}
\newcommand{\ee}{\end{equation}}
\newcommand{\ba}{\begin{eqnarray}}
\newcommand{\ea}{\end{eqnarray}}

\newcommand{\nn}{\nonumber}
\newcommand{\vs}{\vspace{-0.20cm}}

\begin{document}

\begin{center}
\huge{Pion and Kaon Vector Form
 Factors}
\end{center}
\begin{center}
{\large{J. E. Palomar$^{1}$, J. A. Oller$^{2}$, E. Oset$^{1}$}}
\vspace{0.2cm}

\small{$^1$ Departamento de F\'{\i}sica Te\'orica e IFIC, Centro Mixto
Universidad de Valencia-CSIC,\\
46100 Burjassot (Valencia), Spain

$^{2}$ Forschungszentrum J\"ulich, Institut f\"ur Kernphysik (Theorie) \\ 
D-52425 J\"ulich, Germany}
\end{center}
\begin{abstract}{\small{The pion and kaon coupled-channel vector form factors are described
by making use of the resonance chiral Lagrangian results together with a
suitable unitarization method in order to take care of the final state
interactions. A very good reproduction of experimental data is
accomplished for the vector form factors up to $\sqrt{s}\leq 1.2$ GeV and
for the $\pi\pi$ P-wave phase shifts up to $\sqrt{s}\leq 1.5$ GeV.}}
\end{abstract}
\begin{center}
\section{Unitarization}
\end{center}
Using an appropriate unitarization method we take into account the
final state interaction corrections to the tree level amplitudes
 calculated from lowest order $\chi PT$ \cite{GyL} and from the
inclusion of explicit resonance fields in a chiral symmetry fashion as
given in ref. \cite{ecker}. A similar procedure has already been used in the
scalar sector to describe the scalar form factor associated with the
strange-change scalar current $\overline{u}s$ in ref. \cite{jamin}.
Starting from the unitarity of the $S$-matrix and the introduction
of the electromagnetic meson form factor $F_{MM'}(s)$: 
\be
\label{defiff}
\langle\gamma(q)|T|M(p)M'(p')\rangle=e\epsilon_{\mu}(p-p')^{\mu}F_{MM'}(s) 
\ee
with $q^2=s$, $e$ the modulus of the
electron charge and $\epsilon_{\mu}$ the photon
polarization vector, we arrive at the expression: 
\be 
\label{uni1}
\hbox{Im}F_{MM'}(s)=\sum_{\alpha}F^*_{\alpha}(s) \frac{p_\alpha
(s)}{8\pi\sqrt{s}}\theta(s-4m_{\alpha}^2) p_\alpha (s)
\frac{T(s)_{\alpha,MM'}}{p_{MM'}(s)} 
\ee
where $\theta(x)$ is the usual Heaviside function. On the other hand $p_{MM'}(s)$ and $p_\alpha(s)$
 are respectively the moduli of the three momenta of the mesons in the final and
  intermediate meson states, and we sum over intermediate two-meson states.

We will work in the isospin limit, with $|\pi \pi\rangle$ and $|K \bar{K}\rangle$ 
states (and
the $\rho$ resonance) in the $I=1$ channel and only the $|K \bar{K}\rangle$
state (and the
$\omega$ and $\phi$ resonances) in the I=0 channel. Using a matrix
notation, the P-wave amplitudes $T^{I}(s)$ can be written as \cite{N/D}:
\be
\label{Tnd}
T^I=[1+K^I(s)\cdot g(s)]^{-1}\cdot K^I(s)
\ee
where $K^I_{ij}(s)$ are the tree level amplitudes derived from lowest order
$\chi PT$ plus s-channel vector resonance exchange contributions \cite{ecker}
corresponding to the transition $i \rightarrow j$. From eqs. (\ref{uni1}) and
(\ref{Tnd}) it follows, after some algebra, that $F^I(s)$ can be written as:
\be 
\label{ff1} 
F^{I}(s) = \left[1+\widetilde{Q}(s)^{-1}\cdot
K^{I}(s) \cdot \widetilde{Q}(s)\cdot g^{I}(s) \right]^{-1}\cdot
R^{I}(s) 
\ee
where $\widetilde{Q}_{ij}(s) = p_{i}(s)\delta_{ij}$ and
$K^{I}(s)$ is the matrix collecting the tree level amplitudes
between definite $\pi\pi$ and $K\bar{K}$ isospin states.
$F^{I}(s)$ is the column matrix $F^{I}(s)_{i}=F^{I}_{i}(s)$,
$R^{I}(s)$ is a vector made up by functions without any cut and
$g^{I}(s)$ is the diagonal matrix given by the loop with two
meson propagators: 
\be 
\label{g(s)}
g^I_i(s)=\frac{1}{16\,\pi^2}\left[-2+d^I_i+\sigma_{i}(s)\, \log
\frac{\sigma_{i}(s)+1}{\sigma_{i}(s)-1} \right] 
\ee
where $\sigma_{i}(s)=\sqrt{1-4 m_i^2/s}$. We will label pions with $1$ and kaons with $2$ in the $I=1$ case. In the $I=0$
case we only have kaons. 

In the large $N_{c}$ limit loop physics is supressed and then $F^{I}(s)=R_{N_{c}
leading}(s)=F_{t}^{I}(s)$, where $F_{t}^{I}(s)$ is the tree level form
factor.\footnote{We evaluate the tree level form factors and scattering amplitudes using 
the lowest order $\chi PT$ Lagrangian \cite{GyL} plus the chiral resonance 
Lagrangian \cite{ecker}.}

This allows us to write:
\begin{equation}
\label{finalff2} 
F^{I}(s)=\left[1+\widetilde{Q}(s)^{-1}\cdot
K^{I}(s) \cdot \widetilde{Q}(s)\cdot g^{I}(s)\right]^{-1} \cdot
\left[ F_{t}^{I}(s)+R^{I}_{subleading}(s) \right]
\end{equation}
with $R^I_{subleading}(s)$ being of ${\mathcal{O}}(N_c^{-1})$. If we require 
that the vector form factor from eq. (\ref{finalff2}) vanishes for
$s\rightarrow \infty$ as is suggested by the experiments, we find
that the subleading part of $R^{I}(s)$, which at
first can be an arbitrary polynomial (the poles coming
from the resonances are in the leading part $F^{I}_{t}(s)$), must be a constant.
 In order to
fix the constants $R^{I}_{subleading}(s)$ and $d^{I}_{i}(s)$ of
$g^{I}_{i}(s)$ we match our results with those of one loop $\chi
PT$. We take $R^{I=1}_{subleading}(s)=0$ in order to constrain
further our approach. This can be done since we can match our
results with one loop $\chi PT$ by choosing appropriate values
for $d^{I=1}_1$ and $d^{I=1}_2$. The values of the other constants given by the
matching are: 

\ba 
\label{dpi}
d^{I=1}_{1}&=&\frac{m_K^2}{m_K^2-m_\pi^2}\left(\log\frac{m_\pi^2}{\mu^2}+\frac{1}{2}
\log\frac{m_K^2}{\mu^2}+\frac{1}{2}\right)\nn\\
d^{I=1}_{2}&=&\frac{-2\
m_\pi^2}{m_K^2-m_\pi^2}\left(\log\frac{m_\pi^2}{\mu^2}+\frac{1}{2}
\log\frac{m_K^2}{\mu^2}+\frac{1}{2}\right) \nn\\
d^{I=0}&=&\frac{1}{3}+\log\frac{m_K^{2}}{\mu^2}\nn\\
R^{I=0}_{subleading}&=&-\frac{m_K^{2}}{16\sqrt{2}\,\pi^{2}f^{2}}
\left(\frac{1}{3}+\log\frac{m_K^{2}}{\mu^2}\right)
\ea
The bare masses of the resonances
 (which appear in the tree level
quantities) are fixed by the requirements that the moduli of the
$\pi\pi$ $I=1$ and $K\overline{K}$ $I$=0 P-wave amplitudes have a
maximum for $\sqrt{s}=M^{physical}_\rho$ MeV and for
$\sqrt{s}=M_{\phi}^{physical}$ MeV, respectively. For the mass of
the $\omega$ we take directly 782 MeV since there are no
experimental data in the region of the $\omega$ and its
contributions to other physical regions do not depend on such
fine details since the $\omega$ is very narrow. On the other hand, the coupling
of the vector resonances \cite{ecker} to mesons and photons are described by two
real parameters $G_V$ and $F_V$ respectively. We use their experimental value,
$G_V$=53 MeV (from a study of the pion electromagnetic radii \cite{GyL}) and
$F_V$=154 MeV (from the observed decay rate $\Gamma(\rho^{0}\rightarrow
e^{+}e^{-})$ \cite{ecker}).

\section{Results and conclusions.}
As can be seen in fig. 1, we can describe in a very
precise way the vector pion form factor and the P-wave $\pi\pi$ phase
shifts up to about s=1.44 GeV$^2$ (even for higher energies in the case of phase
shifts). For values of $\sqrt{s}$
higher than 1.2 GeV new effects appear: 1) the presence of more
massive resonances, $\rho'$, $\omega'$, $\phi'$... 2) The effect
due to multiparticle states, e.g. $4\pi$, $\omega \pi$ ... which are non
negligible. In figure 2, we compare our results with those of
$\chi PT$. We can see that the resummation of our scheme leads to a much better
agreement with the two loop $\chi PT$ pion vector form factor than with the one
loop one. The resonance regions are also well reproduced.
\begin{figure}[ht]
\centerline{\includegraphics[width=0.9\textwidth]{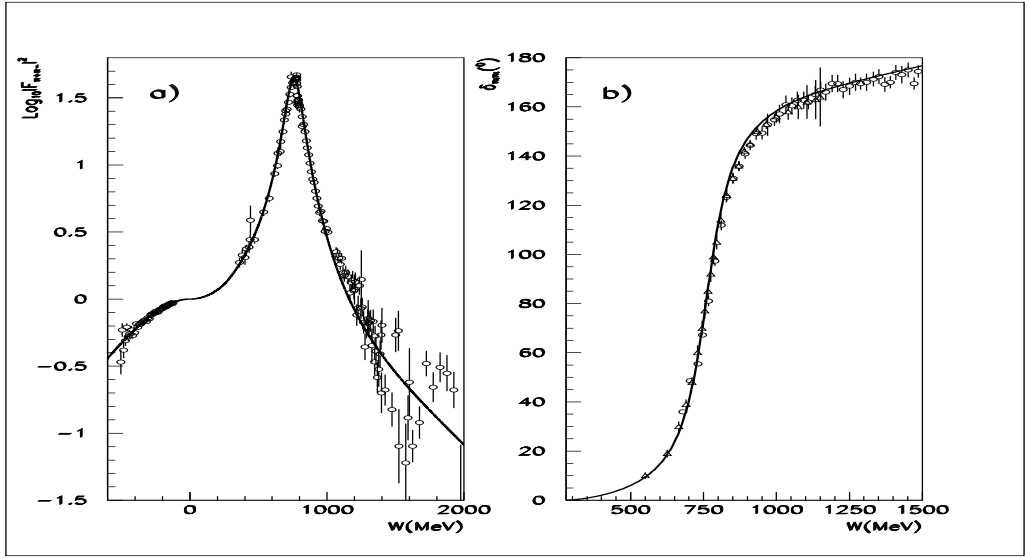}}
\caption{\footnotesize{$W$ is defined as $\sqrt{s}$ for $s>0$ and as
$-\sqrt{-s}$ for $s<0$. a) $\pi^{+} \pi^{-}$
vector form factor. b) $\pi\pi$ P-wave phase
shifts. Both are compared with several experimental data.}}
\end{figure}
\begin{figure}[ht]
\centerline{\includegraphics[width=0.9\textwidth]{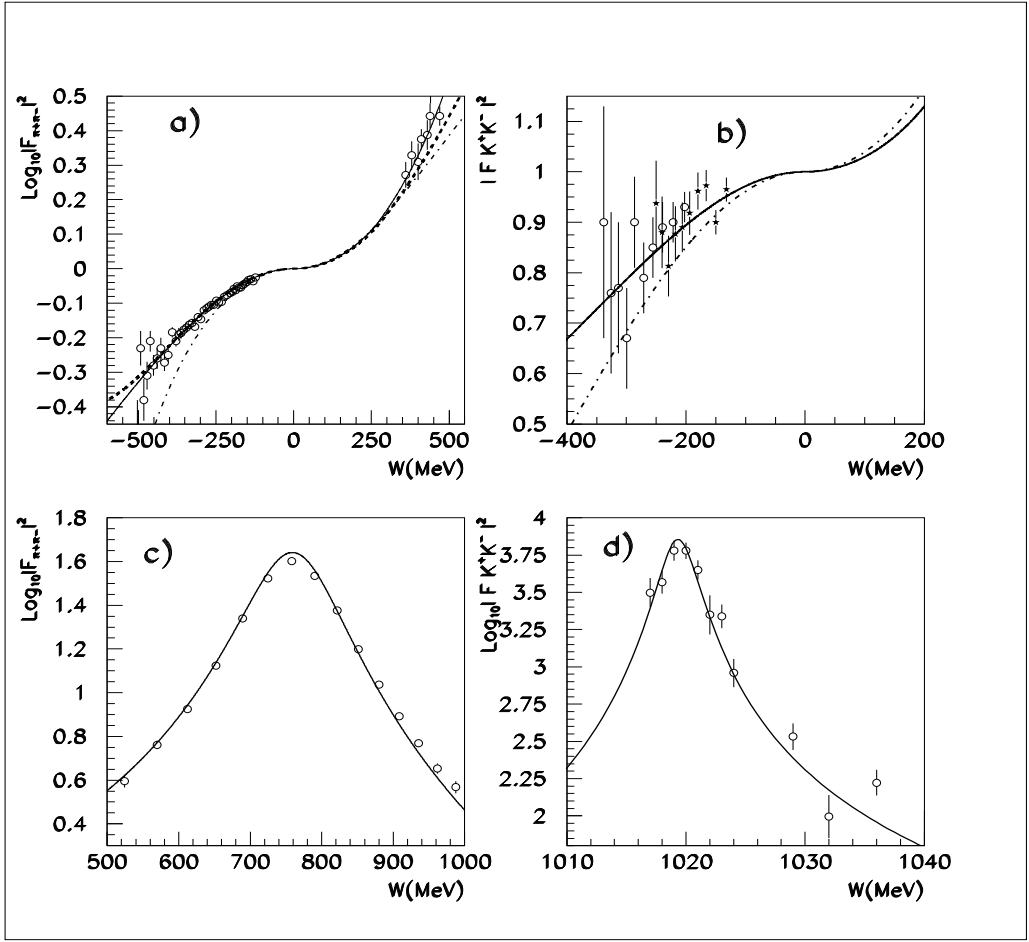}}
\caption{\footnotesize{$W$ is defined as $\sqrt{s}$ for $s>0$ and as $-\sqrt{-s}$ for $s<0$. 
From left to right and top to bottom: a) Vector pion form factor. The dashed-dotted 
line represents one loop $\chi PT$ ref. \cite{GyL} and the dashed one the two loop $\chi PT$ 
result ref. \cite{cola}. b) $K^+ K^-$ form factor. The meaning of the lines is the same as before. 
c)Vector pion form factor in the $\rho$ region. Data from tau decay.
d) $K^+K^-$ form factor. All the results are compared with data.}}
\end{figure}

\end{document}